\documentclass[twocolumn,
3p,
sort&compress
]{elsarticle}
\usepackage[utf8]{inputenc}
\usepackage{graphicx}
\usepackage{xcolor}
\usepackage{comment}
\usepackage{soul}

\usepackage{lineno}

\begin{document}

\title{Single Electron-Hole Pair Sensitive Silicon Detector \\ with Surface Event Discrimination}

\address[1]{Department of Physics \& Astronomy, Northwestern University, Evanston, IL 60208-3112, USA}
\address[2]{Fermi National Accelerator Laboratory, Batavia, IL 60510, USA} 
\address[3]{Department of Physics, University of California, Berkeley, CA 94720, USA}
\address[4]{Department of Physics and Astronomy, and the Mitchell Institute for Fundamental Physics and Astronomy, Texas A\&M University, College Station, TX 77843, USA}
\address[5]{Kavli Institute for Cosmological Physics, University of Chicago, Chicago, IL 60637, USA}
\address[6]{D\'{e}partement de Physique, Universit\'{e} de Montr\'{e}al, Montr\'{e}al, QC  H3T 1J4, Canada}

\author[1]{Ziqing Hong} 
\author[1]{Runze Ren}
\author[2,5]{Noah Kurinsky\corref{cor}}
\ead{kurinsky@fnal.gov} 
\author[1]{Enectali Figueroa-Feliciano} 
\author[6]{Lise Wills}
\author[3]{Suhas Ganjam}
\author[4]{Rupak Mahapatra}
\author[4]{Nader Mirabolfathi}
\author[1]{Brian Nebolsky}
\author[1]{H. Douglas Pinckney}
\author[4]{Mark Platt}

\cortext[cor]{Corresponding Author}


\date{\today}

\begin{abstract}
We demonstrate single electron-hole pair resolution in a single-sided, contact-free 1~cm$^2$ by 1~mm thick Si crystal operated at 48~mK, with a baseline energy resolution of 3~eV. This crystal can be operated at voltages in excess of $\pm50$~V, resulting in a measured charge resolution of 0.06 electron-hole pairs. The high aluminum coverage ($\sim$70\%) of this device allows for the discrimination of surface events and separation of events occurring near the center of the detector from those near the edge. We use this discrimination ability to show that non-quantized dark events seen in previous detectors of a similar design are likely dominated by charge leakage along the sidewall of the device.
\end{abstract}

\maketitle

\section{Introduction}

Research into cryogenic calorimeters with eV-scale energy thresholds has grown in recent years, driven in large part by the needs of low background physics experiments, in particular direct detection of sub-GeV dark matter (DM) and coherent neutrino scattering measurements (CE$\nu$NS) \cite[and references therein]{SingleEH,HVeV,Nucleus,CRESSTsurface,edelweiss,coherent}. The recent demonstration of single electron-hole pair resolution in a cryogenic silicon crystal showed that the Neganov-Trofimov-Luke (NTL) effect~\cite{Luke,Neganov} can be leveraged to amplify the initial recoil energy by applying a bias voltage across the sensitive volume. For electron recoil events, this amplification is not quenched, and thus turns a calorimeter into a charge amplifier with single charge resolution~\cite{SingleEH}. This means that a single detector can operate both as a highly sensitive eV-scale calorimeter with 0~V bias voltage, suitable for applications like nuclear-recoil detection (including CE$\nu$NS searches~\cite{Nucleus}), and complement high-resolution CCDs~\cite{skipper,SENSEI} with phonon energy information when run in single-charge sensitive mode with NTL gain. Here we focus on the charge detection aspect of these detectors for rare event searches; for a more detailed exploration of cryogenic detectors applied to nuclear-recoil searches, and CE$\nu$NS in particular, we refer the readers to Ref~\cite{Nucleus}.

In the context of a rare event search, the optimal detector design will minimize both charge and energy resolution, and at the same time not introduce excessive backgrounds. Use of the NTL effect produces an additional low-energy background from `dark counts'~\cite{SingleEH,HVeV}, which can be produced by mechanisms including charge leakage through the interfaces between the electrodes and the bulk (`interface leakage') and through generation of unpaired excitations in the detector bulk, the electrode surfaces, or the `outer' non-instrumented surfaces. These dark counts are the currently dominant background of electron-recoil dark matter searches with this type of detector~\cite{HVeV}.

Depending on the production mechanism, different strategies to minimize this background can be utilized. Breaking the degeneracy of different causes of dark counts is a crucial step for improving the scientific reach of the NTL-effect-driven detectors. As an examples, if the process depends on electric field strength, a better energy resolution allows for lower field strength to be used to attain the same charge resolution, which allows for a reduction in dark counts. If the process occurs preferentially in particular surfaces, devices with good position resolution can reduce dark counts through fiducial volume cuts. Finally, the contributions from interface leakage could be reduced by using a contact-free biasing scheme, which does away with the electrode/surface interface on one side of the device.

In this paper, we present a detector with improved energy and position resolution compared to that discussed in Ref.~\cite{SingleEH}, explore the impact of contact-free operation on dark counts, and take advantage of the good position dependence of this detector to study the origin of the dark counts.


\begin{figure}[htbp]
    \centering
    \includegraphics[width=\columnwidth]{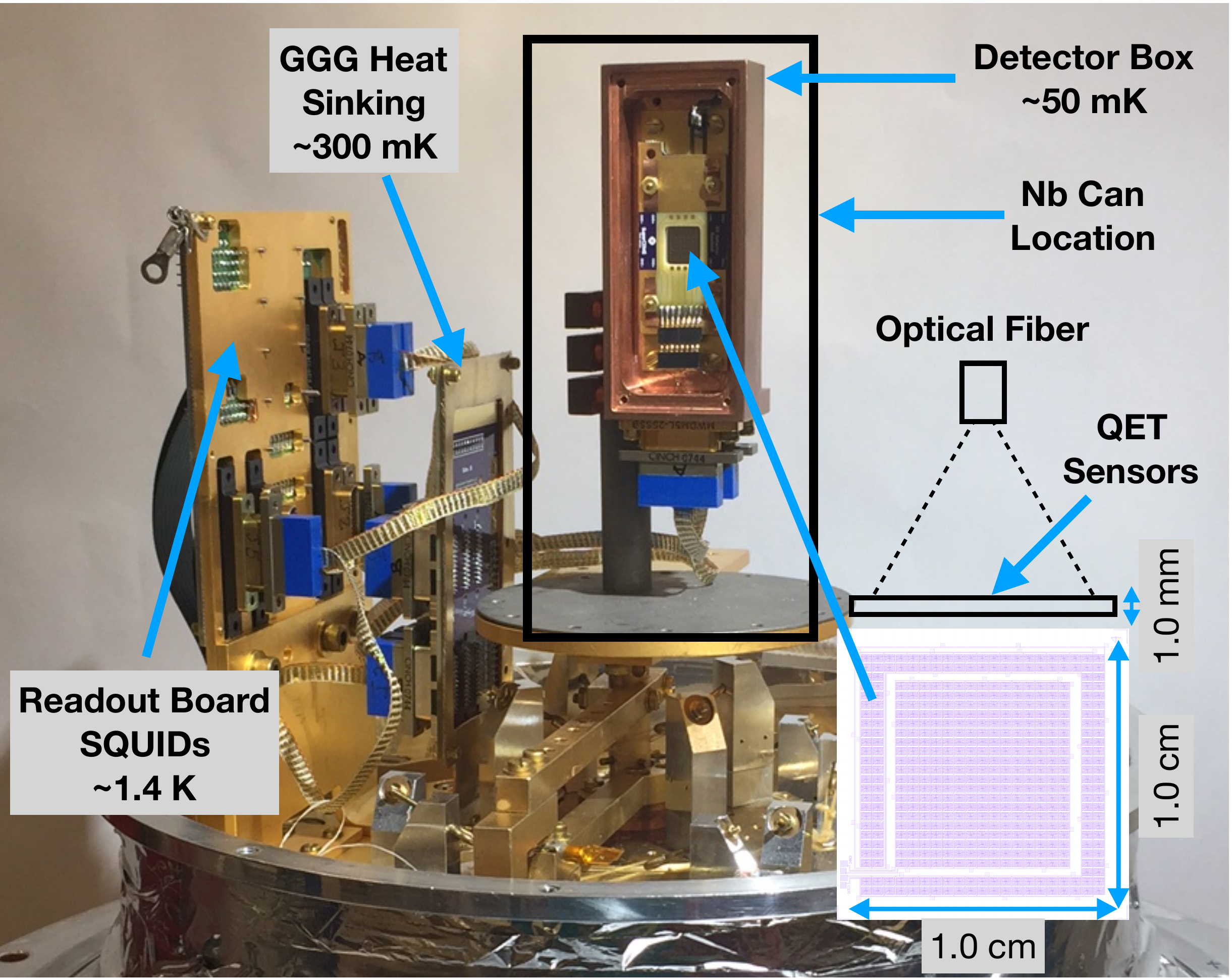}\\    
    \caption{A side view of the detector box mounted inside the ADR with the outer shielding removed. The inset picture shows the schematics of the detector used, together with the optical fiber and its field of illumination. The cartoon shows the detector and laser from the side; the detailed diagram of the two phonon channels is a top-down view.}
    \label{fig:setup}
\end{figure}

\section{Experimental Setup}

The data described in this paper were taken in a VeriCold Adiabatic Demagnetization Refrigerator (ADR), cooled to 48~mK. We fabricated a silicon detector $1~\mathrm{cm^2}\times  1~\mathrm{mm}$ in size. The bottom surface was polished but uninstrumented. The top surface of the detector was instrumented with Quasiparticle-trap-assisted Electrothermal-feedback Transition-edge sensors (QETs) for phonon measurement. Each QET consists of a set of aluminum fins that absorb phonons and concentrate their energy into the transition edge sensor (TES), which acts as a high-gain power to current amplifier. 

The QETs were arranged into inner and outer channels, as illustrated in Fig.~\ref{fig:setup}. The detector readout scheme is the same as that described in Ref~\cite{SingleEH}, with DC superconducting quantum interference devices (SQUIDs) operated in closed-loop mode. 300/168 QETs with critical temperature ($\mathrm{T_c}$) $\sim63$~mK in the inner/outer channel are connected in parallel, then in parallel with a 50~m$\mathrm{\Omega}$ shunt resistor at 48~mK, which in turn is connected to a current bias circuit at room temperature. The design of the QETs was the same as for the first device described in Ref~\cite{SingleEH}, but we increased the total phonon absorber coverage on the top surface from 25\% to 70\% to enhance the phonon absorption rate as well as position dependence of the phonon signal.

The detector was clamped between two printed circuit boards (PCBs). The top PCB hosts the electrical readout contacts for the QETs and a grounded copper plane around the device. The bottom PCB was coated with a $4~\mathrm{cm^2}$ copper square with the detector placed in the center. The copper square served as the high-voltage (HV) electrode. Four small pieces of cigarette paper $\sim$13~$\mu$m thick were placed under the four corners of the detector to insulate the silicon crystal from the electrode. The vacuum gap between the silicon and the electrode depends on the thickness of the cigarette paper under a given amount of pressure at 48~mK, thus the voltage across the crystal needs to be calibrated. 

During operation, the HV electrode was voltage biased, while the `ground' of the QET circuit was held at 0~V. Due to limitation of the electronics, the highest crystal bias for stable operation is limited to below 160~V across the electrodes, corresponding to 50~V across the crystal with the calibration detailed in Sec.~3.  This setup allowed for a nearly homogeneous electric field inside the silicon crystal. The detector assembly was placed in, and heat sunk to, a copper box that was designed to be light tight. The copper box was mounted on the base temperature stage of the ADR. A superconducting Niobium enclosure surrounds the copper box, serving as a magnetic shield. 

For calibration purposes, we employ two photon feedthrough systems for optical photons and soft X-rays. First, a plastic optical fiber with a core diameter of 1~mm was fed through the detector box, with the gap between the fiber and the box filled with Eccosorb epoxy~\cite{Eccosorb}. The plastic optical fiber was coupled to a single-mode optical fiber~\cite{SM450} through two pieces of KG-3 glass at 1.4~K. The single-mode fiber and the KG-3 glass filter were chosen to attenuate infrared photons from ambient and black body radiation from higher temperature stages. The other end of the single mode fiber was connected to a vacuum feed-through at room temperature, then to a laser diode with a wavelength of 635~nm~(corresponding to 1.95~eV per photon)\cite{LPS-635}. For the second feedthrough system for soft X-rays, a $1~\mathrm{cm^2}$ square opening was cut on the copper box lid and re-sealed with a piece of aluminum foil 0.17~mm thick. The opening aligned with a Beryllium window installed on the ADR, serving as an X-ray input port. Multiple layers of Aluminized mylar sheets were placed between the opening and the Beryllium window at different thermal stages to block black body radiation from higher temperature stages while presenting minimal X-ray attenuation.

\begin{figure}[htbp]
    \centering
    \includegraphics[width=\columnwidth]{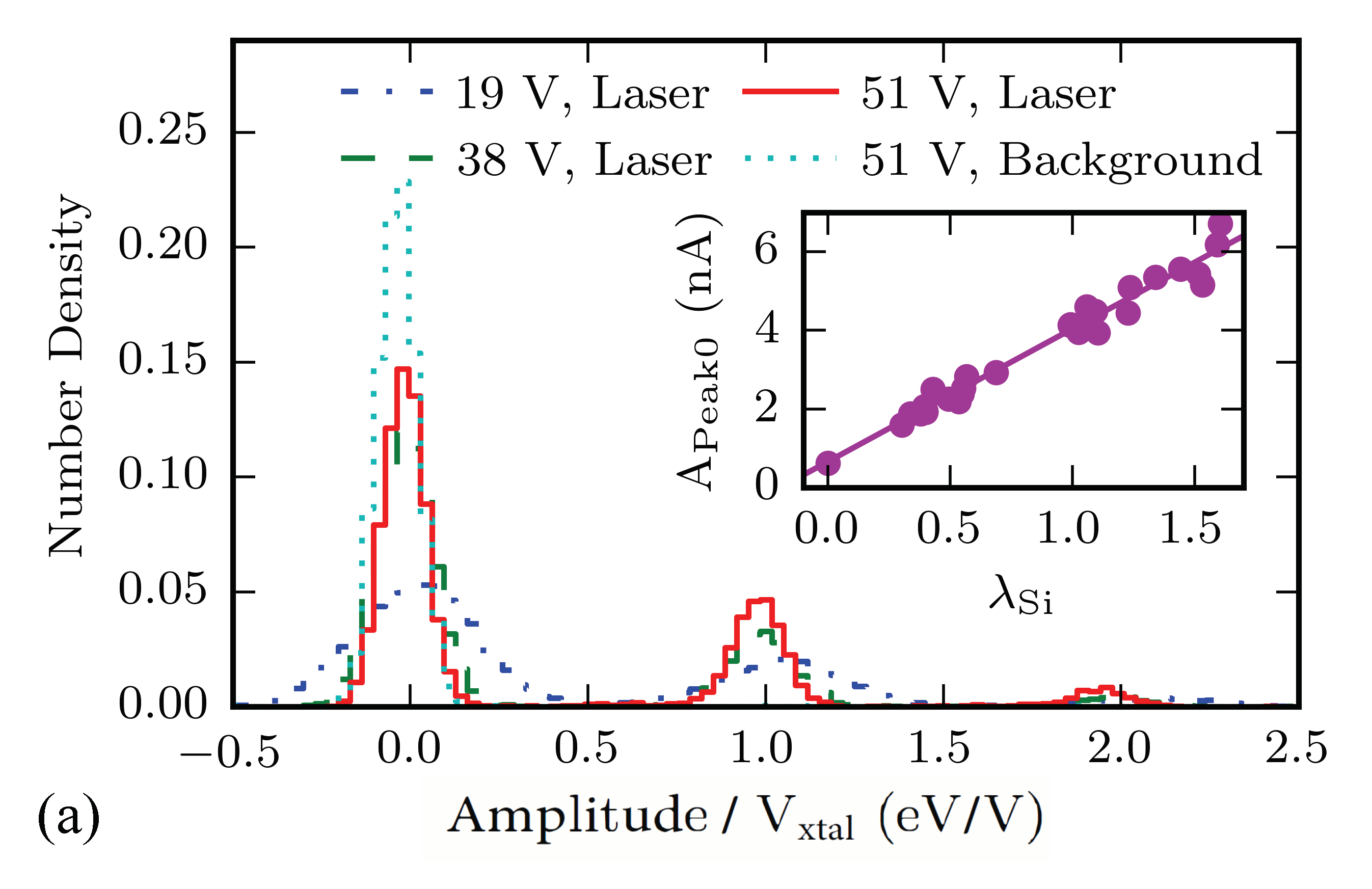}
    \includegraphics[width=\columnwidth]{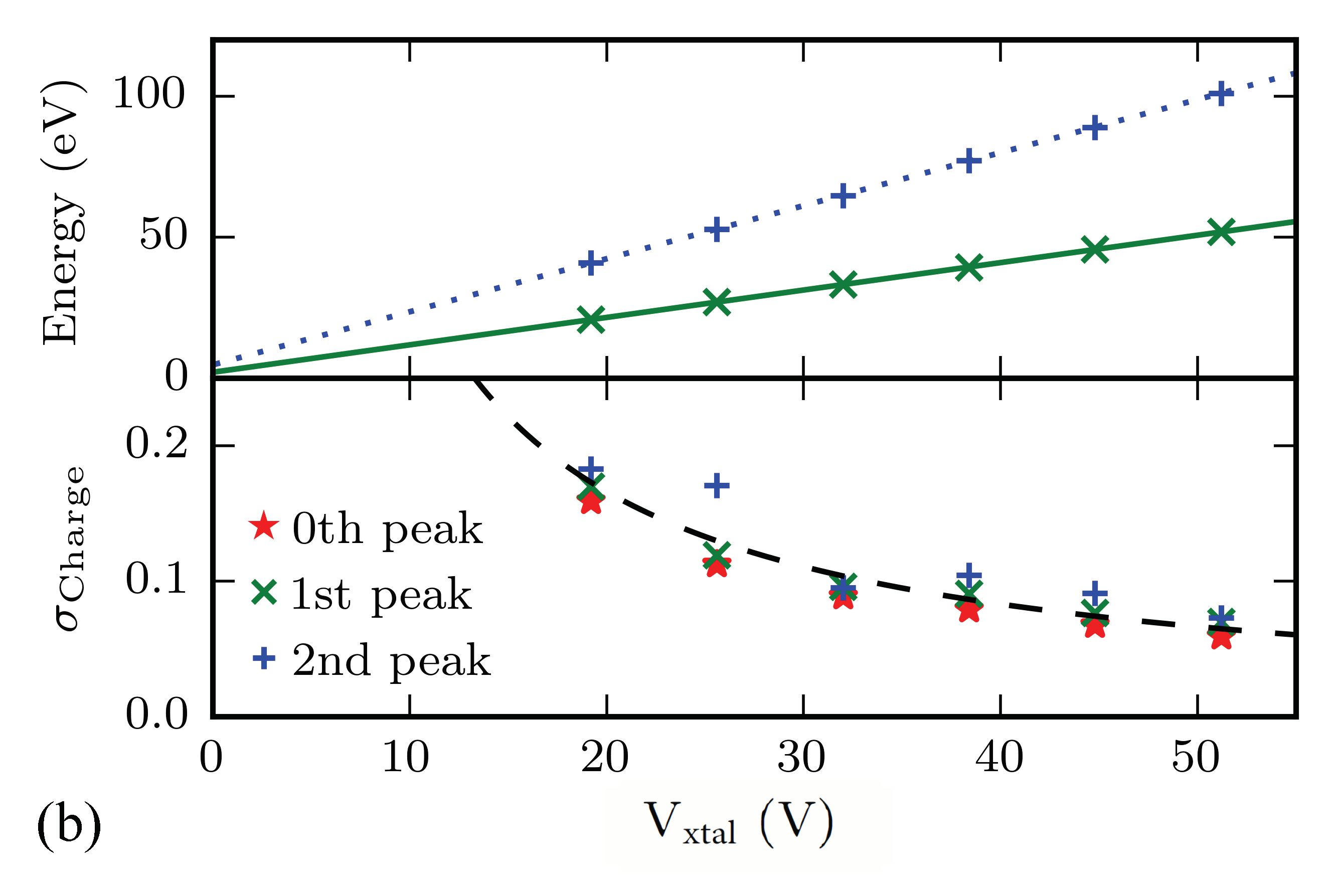}
    \caption{(a): Voltage-normalized laser spectra as a function of crystal bias voltage ($V_{\mathrm{xtal}}$), showing the signal to noise improving and the reduced effect of prompt phonons as bias is increased. The signal peaks become narrower with increased bias, and the offset from the number of charge carriers decreases. The inset shows the dependence of the zero electron-hole pair offset on the measured mean photon number, indicative of photon absorption in the QETs independent of that in the detector bulk. (b): (Top panel) Measured energy deposition in the silicon bulk for the 1 and 2 electron hole pair peaks as a function of crystal bias voltage, after correcting for the surface energy depositions. (Bottom Panel) Charge resolution as a function of crystal bias voltage, measured from each discrete peak. The zero-th peak points show the baseline charge resolution. The charge resolution very closely follows the ideal scaling of $V_{\mathrm{xtal}}^{-1}$ shown as a dashed line; this corresponds to a linear increase in signal and no increase in noise as a function of voltage, demonstrating that the noise is insensitive to these voltages.}
    \label{fig:laser}
\end{figure}


\section{Energy and Charge Resolution}

To calibrate the voltage drop across the crystal, establish an absolute energy scale of the signal, and measure the phonon energy resolution, the laser was pulsed with a fixed width of 500~ns and a frequency of $\sim$100~Hz. The readout was triggered on the laser driver signal in order to read out zero-photon events, which are nominally below the threshold of the detector. Multiple data sets were acquired at different crystal bias voltages and laser intensities. The Fig.~\ref{fig:laser}a shows the laser calibration spectra in units of electron-hole pairs with a few example voltages applied across the crystal. The calibration of the voltage will be described in the following paragraphs. 

Due to the $\sim$70\% overall QET coverage on the instrumented surface, and $\sim$90\% coverage near the center of the laser spot, significant photon energy was being absorbed by the phonon sensors before the photons could reach the detector bulk, producing no electron-hole pairs to mediate NTL gain. For a constant laser intensity, an event has $n$ photons absorbed in the bulk, following a Poisson distribution with a mean of $\lambda_{\mathrm{Si}}$, and a mean of $\lambda_{\mathrm{s}}$ photons absorbed directly by the QETs on the surface. The total energy measured by the QETs for such an event is given by
\begin{equation}
E_{\mathrm{QET}}(n)=\epsilon_{\mathrm{ph}}n\left(E_{\gamma}+f_\mathrm{xtal} \cdot q_e V_{\mathrm{app}}\right)+\epsilon_{\mathrm{s}}E_{\gamma}\lambda_{\mathrm{s}},\label{eq:diffspectrum}
\end{equation}
where $\epsilon_{\mathrm{s}}$ is the energy efficiency of surface events, $\epsilon_{\mathrm{ph}}$ is the energy efficiency for events in the bulk (phonon energy efficiency), $E_{\gamma}=1.95$~eV is the photon energy, $q_e$ is the electron charge, $V_\mathrm{app}$ is the voltage across the electrodes, and $f_\mathrm{xtal}$ is the fraction of $V_\mathrm{app}$ across the crystal. We refer to $V_\mathrm{xtal} = f_\mathrm{xtal}\cdot V_\mathrm{app}$ as crystal bias voltage. The rest of $V_\mathrm{app}$ goes across the vacuum gap. We note that $\epsilon_{\mathrm{s}}$ is not affected by $V_{\mathrm{app}}$ as the QETs are held at ground potential, while $\epsilon_{\mathrm{ph}}$ is also independent of $V_{\mathrm{app}}$ within 1\%~\cite{SingleEH}.

Due to quantized number of electron-hole pairs produced, for sub-charge resolution, we can measure the mean number of photons absorbed in the bulk, $\lambda_{\mathrm{Si}}$, by comparing the number of events under discrete electron-hole-pair peaks with a Poisson distribution.
As shown in Fig.~\ref{fig:laser}a, the offset of the zero-photon peak ($n=0$), which serves as a measurement of the surface absorption ($\epsilon_{\mathrm{s}}E_{\gamma}\lambda_{\mathrm{s}}$), is proportional to $\lambda_{\mathrm{Si}}$, and thus is linear in laser power and $\lambda_{\mathrm{s}}$. This allowed us to correct the measured energy scale for a given laser power, removing the average energy from surface energy depositions. Additional variance due to these surface events persists. For this reason, a low laser intensity of $\lambda_\mathrm{Si} \sim 0.4$ is used for the resolution studies.

After correcting for surface energy depositions, we observed the bulk event energy as a function of voltage as shown in the upper panel of Fig.~2b. We extrapolate the linear relation with $n=1, 2$ to $V_{\mathrm{xtal}} = 0~\mathrm{V}$. The intercepts correspond to energy depositions of $1.95~\mathrm{eV}$ and $3.90~\mathrm{eV}$, respectively. We use these to calibrate the energy scale of this detector. The slopes of the linear relations are used to calibrate $f_\mathrm{xtal}$. For this contact-free mounting scheme, the voltage across the crystal varied from 30\% to 45\% of the applied bias for different mounting techniques, but for a given run we find that this fraction is stable as long as charge buildup is mitigated.

An important detector performance parameter for the cryogenic calorimeters is the phonon energy efficiency. With the calibration of this detector, we inferred a phonon energy efficiency of $\epsilon_\mathrm{ph}\sim 27\%$, with a 95\% confidence interval of 22\% to 30\%. This broad uncertainty is due to systematic uncertainties on the resistance values in the readout circuit, as well as uncertainties in the crystal bias calibration and surface absorption correction. The energy efficiency is significantly higher than the $\sim$4\% efficiency measured for the previous device described in Ref.~\cite{SingleEH}. This is potentially due to the high aluminum coverage (70\% as compared to 25\%) leading to more efficient phonon collection, and the fact that this device is instrumented on only one side, while the back side of the crystal is left bare, acting as an athermal phonon reflector rather than a phonon sink. 


We measure a baseline phonon resolution of 3.0$\pm 0.5$~eV from the width of the background peak in Fig.~\ref{fig:laser}a.
This phonon resolution is 4 times better than that measured in Refs.~\cite{SingleEH, HVeV} due to the much higher energy efficiency, despite the fact that the $\mathrm{T_c}$ of this device was 12~mK higher. With more than 20~V across the crystal, we can resolve individual peaks at 99\% confidence. At 50~V, we obtain a charge resolution of $\sim0.06$~$e^{-}h^{+}$ pairs, which is comparable to the charge resolution obtained in Ref~\cite{HVeV} but at a much lower voltage. The charge resolution would likely improve at higher voltages, but we were limited to 50~V by constraints in the electronics as discussed earlier.

\begin{figure}[tbp]
    \centering
    \includegraphics[width=\columnwidth]{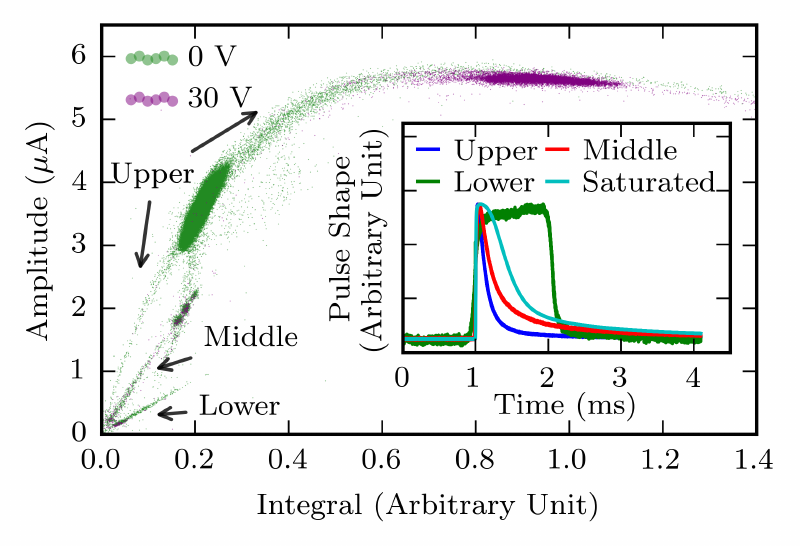}\\

    \caption{Pulse amplitude as a function of pulse integral at 0~V and 30~V from $^{55}$Fe radioactive source. This demonstrates that only the events in the upper track scale with voltage. The middle population of events are consistent with surface hits on the QETs. The insert shows that the lower population are square ``glitch" events, likely caused by transient RF power spikes. A cut in this plane allows surface event rejection by pulse shape discrimination. See text for discussion.}
    \label{fig:integral}
\end{figure}

\section{Surface Event Reconstruction}

In order to calibrate the detector over a larger energy scale, and to probe position-dependent effects on our energy reconstruction, data were acquired using two sources. The first was a $^{55}$Fe source with two prominent X-ray lines at 5.9 and 6.4~keV, and the second was a $^{57}$Co source with a prominent 122~keV line. Due to the vacuum-gap design, this device is prone to charge buildup when subjected to the large charge production rate induced by the $^{57}$Co X-ray source. As electron-hole pairs are generated, they accumulate at the insulated surface, resulting in a counter voltage built across the vacuum gap that reduces the voltage across the crystal. A 0.2~V/hr voltage gain decrease was observed in $^{57}$Co data at an event rate of $\sim$ 1 Hz, consistent with the expected charge generated by this event rate. For a voltage bias of 50~V, this corresponds to a 0.4\% degradation in energy resolution per hour, and can be corrected by interspersed laser calibration data as in Ref~\cite{HVeV}. Grounding the HV electrode and warming up the detector to $>$20~K was found to neutralize this built-in potential, while grounding at 4~K was not always sufficient. Other neutralization methods are being investigated; in particular, we are studying how neutralization state and biasing history can affect the dark event rate. These studies will be discussed in a future work.

X-rays with energies below 100~keV have a mean free path much less than 1~mm in Si and Al. The $^{55}$Fe X-rays are therefore predominantly absorbed on the surface of the detector. Given that the sources face the instrumented side of the detector, this produces a large population of $^{55}$Fe X-ray hits on the QETs, rather than in the detector bulk. Fig.~\ref{fig:integral} shows the QET pulse integral compared to the pulse height obtained using an optimal filter\cite{Kurinsky}. While the pulse height is proportional to the energy in the small signal limit, the pulse duration begins to lengthen with smaller changes in pulse amplitude when the QETs approach the saturation regime, as shown in the insert of Fig.~\ref{fig:integral}. This produces the flat portion of this curve, where the pulse amplitude is only weakly dependent on the pulse integral. The proportionality between pulse height and integral therefore depends on the pulse shape. Fig.~\ref{fig:integral} shows that there are 3 distinct lines in the small-signal region of different proportionality. Of these lines, only the upper one scales with voltage, which indicates that these are bulk events. Upon inspection, the lowest track is a population of square pulses generated by out of band RF pickup, which appears in the QET as a time-dependent and abrupt change in bias power. 

The middle class of events appear to be real QET events, but have a long secondary tail; the fact that they do not scale with applied voltage is consistent with these events occurring in or very close to the QETs, and are thus surface events. This event shape is well understood as a primary event in the aluminum fin which emits phonons back into the substrate; the primary event is seen in a single QET, thus heavily saturates it, while the phonons are seen on a longer timescale in the adjacent QETs, producing a pulse with two falltimes. We can therefore remove this population with high efficiency even at single electron-hole pair energies by a selection criterion in this integral versus pulse amplitude space, taking advantage of pulse-shape information. There is a small band of events that extend between the middle and upper event classes that are likely near-surface events with reduced charge production and partial absorption of initial event energy by the nearest QETs. The integral cut also removes the majority of these events.

\begin{figure}[t]
    \centering
    \includegraphics[width=\columnwidth]{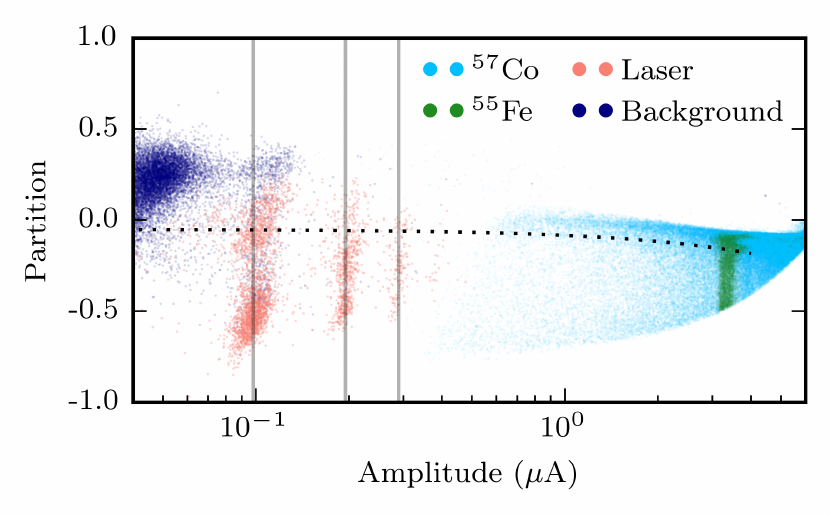}
    \caption{Radial partition, or the relative difference in energy absorbed by each channel, as a function of pulse height. Data shown from $^{57}$Co and $^{55}$Fe sources with no bias across the crystal, as well as laser and background (no source) data with a crystal bias of 50~V. A partition near +/-1 indicates an energy deposition entirely in the outer/inner channel, respectively. All events shown are those that pass the pulse-shape cut to remove surface events, described in the text. For events near the center of the detector (negative partition) the laser and background events are quantized, while non-quantized background events are restricted to the outer part of the detector. The dashed line represents the 50\% efficiency cut separating inner from outer events, calibrated using the $^{57}$Co data.}
    \label{fig:partition}
\end{figure}


The 122~keV X-rays of the $^{57}$Co, unlike the 6~keV from the $^{55}$Fe, have a much longer mean free path in Si, and are more likely to Compton scatter than be absorbed in our thin Si substrate. This means that the $^{57}$Co events are primarily distributed uniformly in energy and position within the detector, with a range of energies between 0 and 50~keV. This provides us a means with which to study the energy partition between the inner and outer channels as a function of event energy. Fig.~\ref{fig:partition} shows the $^{57}$Co and $^{55}$Fe data in partition space for data taken at 0~V, along with laser and background events acquired at a crystal bias of 50~V. A higher trigger threshold was used for the $^{57}$Co data to avoid the excessive trigger rate caused by the long-lived thermal tails from the very high energy events. This resulted in the cutoff for the $^{57}$Co data around 0.4~$\mu$A. The non-vertical feature was caused by the trigger set only on the inner channel. Above this energy, the $^{57}$Co data demonstrates that the $^{55}$Fe events occur across the face of the crystal, and that by employing the pulse shape selection described earlier, the remaining events fill a single continuous band across the partition space as expected.

\section{Discussion}

With the high QET coverage and a thin silicon substrate, this device exhibits enhanced partition resolution over past devices. This sheds light on the origin of the dark counts in the contact-free biasing scheme. As shown in Fig.~\ref{fig:partition}, in the small-signal region, we extrapolated the partition down to threshold of this detector. This shows that the laser events filled out a wide range of the partition space, but were biased towards the inner channel. This is due to the laser pointing towards the center of the detector, as shown in the insert of Fig.~\ref{fig:setup}. We found that the background events that are non-quantized are mostly contained near the outer channel. We used the $^{57}$Co data to construct an approximately 50\% efficient radial partition, shown by the dotted line in Fig.~\ref{fig:partition}. By rejecting events larger than this partition requirement, we reject 95\% of non-quantized dark events and 80\% of quantized dark events, while keeping 90\% of laser events. We observed very few events above one electron-hole pair in the inner region of the detector. This suggests that, for this contact-free design, charge leakage towards the center of the detector is dominated by surface physics, and that the majority of the non-quantized events are occurring along the sidewall of the device. Qualitatively, Figure~\ref{fig:partition} shows that, while the laser events can have inner-like and outer-like partitions, the background is primarily outer-like, such that a partition cut can be used to reject the majority of the charge leakage events.


We also note that the surface rejection demonstrated by Fig.~\ref{fig:integral} rejects surface events on the instrumented detector surface, but will not reject surface events on the opposite surface. A detector design which could benefit by more than a factor of 2 in background rejection would need a two-sided readout, such that this method could be applied to both detector faces. Operating two-sided detectors, and solving surface dark counts, are currently orthogonal directions of research, but as sources of dark counts are better understood on single-sided detectors, subsequent work on a two-sided readout scheme will allow for much larger rejection of external backgrounds in a future large-scale detector payload.

\section*{Acknowledgements}
ZH and RR are supported by NSF Grant PHY-1809730. This document was prepared by NK using the resources of the Fermi National Accelerator Laboratory (Fermilab), a U.S. Department of Energy, Office of Science, HEP User Facility. Fermilab is managed by Fermi Research Alliance, LLC (FRA), acting under Contract No. DE-AC02-07CH11359. We thank Matt Pyle for the mask design for this device and discussions thereof, Blas Cabrera and Martin Huber for support in the electronics readout, Noemie Bastidon for her work in the preliminary design of our optical fiber setup and wirebonding, and SLAC for making available their computing resources.

\bibliographystyle{elsarticle-num}
\bibliography{refs}

\end{document}